# Spin-like current from phase space distributions

Peter Holland

Green Templeton College
University of Oxford
Woodstock Road
Oxford OX2 6HG
England

peter.holland@gtc.ox.ac.uk

04.01.2009

## Abstract

The spin 0 generalized phase space approach provides a general expression for local current which depends on the choice of distribution function and generally deviates from the Schrödinger current. It is shown that the continuity equation restricts the admissible bilinear distributions such that the current has a unique dependence on the wavefunction and coincides with the non-relativistic limit of the relativistic spin $\frac{1}{2}$ current for a spin eigenstate, up to a constant vector. Examples of non-bilinear distributions that have the latter property are given.

## 1. Introduction

It is well known that the current implied by the Schrödinger equation is not unique, being defined only up to a divergence-free addition. Some purchase on this problem was gained by the observation that a unique expression for the 4-current of a spin $\frac{1}{2}$ particle is enforced by Lorentz covariance [1]. In the non-relativistic limit, and assuming a spin eigenstate and no external magnetic field, the resulting expressions for the density and 3-current are

$$\rho = \psi^*\psi, \quad j_i = \frac{\hbar}{2mi}\left(\psi^*\frac{\partial\psi}{\partial q_i} - \frac{\partial\psi^*}{\partial q_i}\psi\right) + \frac{\hbar}{2m}\varepsilon_{ijk}\frac{\partial\rho}{\partial q_j}a_k, \quad i,j,k = 1,2,3, \qquad (1.1)$$

where $\psi$ obeys the spin 0 Schrödinger equation and $(\hbar/2)a_i$ is the spin vector, $a_i$ being a constant unit vector. Thus, even though it plays no role in the dynamics, the current contains a contribution from the spin (the 'spin current') in addition to the usual expression quoted for the current implied by the Schrödinger equation. Some implications of the modified expression (1.1) have been explored [2-4].

The question raised here is whether there is an explanation for the additional term and its uniqueness already in non-relativistic spin 0 theory, without going to relativity or directly invoking spin. That there may be such an anticipation follows since, of the infinite number of divergence-free vectors we may add to the Schrödinger current, the spin current is clearly one, the only feature foreign to the Schrödinger equation being the direction $a_i$. To gain an alternative perspective, we examine this question within the generalized (quasi-)distribution approach in which Hilbert space operators are replaced by phase space $c$-numbers, the correspondence being characterized by the choice of distribution function. The value of this formalism in the present context is that it provides a formula to generate the set of all local currents, each distinguished by the distribution. The condition that the current obeys the continuity equation in accordance with the conservation of $\rho$ restricts the distribution functions; we call these 'admissible'. It turns out that for the admissible distribution functions that are bilinear combinations of the wavefunction the current uniquely implied by this method is just (1.1) with $a_i$ in the spin term replaced by an arbitrary constant vector (the spin term in this case will be called 'spin-like'). This class of functions, which apparently has not been considered previously, extends to some non-bilinear distributions and the relevant distributions need not have the quantal momentum density as a marginal. The current obtained uniquely as a residue of relativistic spin $\frac{1}{2}$ theory is therefore also obtained, up to a constant vector, from a quite different starting point in non-relativistic spin 0 theory.

**2. Spin-like current from bilinear distributions**

A general expression for the phase space distribution that yields the quantal position and momentum densities as marginals is given by [5]

$$F(q,p) = (1/2\pi)^6 \int \exp\left[-i\tau_i p_i + i\theta_i(u_i - q_i)\right] f(\theta,\tau) \psi^*\left(u - \tfrac{1}{2}\hbar\tau\right) \psi\left(u + \tfrac{1}{2}\hbar\tau\right) du\, d\tau\, d\theta \quad (2.1)$$

where

$$f(\theta, 0) = 1 \qquad (2.2a)$$

$$f(0, \tau) = 1. \qquad (2.2b)$$

The condition (2.2a(b)) ensures that the position (momentum) distribution is obtained by integration over momentum (position). We always require (2.2a) but (2.2b) is not essential in our considerations and relaxing it yields an even wider class of distributions. The distribution is characterized by the kernel $f(\theta,\tau)$ (for examples see [6, 7]) which may depend on $\psi$ and so this set of functions goes considerably beyond the bilinear distributions of which the Wigner function is the most famous example. The latter is obtained when $f(\theta,\tau) = 1$:

$$W(q,p) = (1/2\pi)^3 \int \exp(-i\tau_i p_i)\psi^*(q - \tfrac{1}{2}\hbar\tau)\psi(q + \tfrac{1}{2}\hbar\tau)d\tau. \tag{2.3}$$

Insight into the expression (2.1) may be obtained by noting that it can be written as a smeared Wigner function [8]:

$$F(q,p) = \int g(q-q', p-p')W(q',p')dq'dp' \tag{2.4}$$

where

$$g(q,p) = (1/2\pi)^6 \int \exp[-i\tau_i p_i - i\theta_i q_i] f(\theta,\tau)d\tau d\theta. \tag{2.5}$$

It was suggested by Cohen [9] that one can generate all possible expressions for the local kinetic energy by adopting the classical expression for the mean energy using the phase space function (2.1) as a weight. Here we employ a similar approach to obtain all local currents. Thus, defining

$$j_i(q) = (1/m)\int p_i F(q,p)dp, \tag{2.6}$$

we obtain from (2.1)

$$j_i = \frac{\hbar}{2mi}\left(\psi^* \frac{\partial \psi}{\partial q_i} - \frac{\partial \psi^*}{\partial q_i}\psi\right) - \frac{i}{m}\left(\frac{1}{2\pi}\right)^3 \int \exp\left[i\theta_j(u_j - q_j)\right]\frac{\partial f(\theta,\tau)}{\partial \tau_i}\bigg|_{\tau=0} \rho(u)du d\theta. \tag{2.7}$$

where (2.2a) (and not (2.2b)) has been used. We therefore deduce the usual Schrödinger current and an additional term characterized by the choice of *f*. This expression reduces to the Schrödinger current when

$$\frac{\partial f(\theta,\tau)}{\partial \tau_i}\bigg|_{\tau=0} = 0, \tag{2.8}$$

a condition obeyed by an infinite class of kernels [10]. We are interested in the general case where the kernel does not satisfy (2.8) and consider the following three constraints:

(a) In order that the total current corresponds to the density $\rho$, the continuity equation implies that the added term must be divergence-free:

$$\int \theta_i \exp\left[i\theta_j(u_j - q_j)\right]\frac{\partial f(\theta,\tau)}{\partial \tau_i}\bigg|_{\tau=0} \rho(u)du d\theta = 0. \tag{2.9}$$

This condition, which must hold for all $\psi$, supplies a restriction on the admissible $f$s that apparently has not been considered before (and provides an additional criterion against which to judge the various choices made for $f$ in the literature).

(b) We require that the current is real. From (2.6) this follows if $F$ is real which is guaranteed when $f^*(\theta,\tau) = f(-\theta,-\tau)$ [5].

(c) Although not essential to our analysis, it is usual to require that the mean momentum implied by the current equals the quantum expression. We have for (2.7)

$$m \int j_i(q) dq = \langle \hat{p}_i \rangle - i \left. \frac{\partial f(0,\tau)}{\partial \tau_i} \right|_{\tau=0} \tag{2.10}$$

and hence the desired equality is obtained when the $f$-term on the right-hand side vanishes. This requirement is satisfied in all the examples we give below; (2.2b) is sufficient to achieve it but, as we shall see, not necessary.

Restricting to bilinear distributions, so that $f$ is independent of $\psi$, (2.9) entails

$$\theta_i \left. \frac{\partial f(\theta,\tau)}{\partial \tau_i} \right|_{\tau=0} = 0 \tag{2.11}$$

and hence

$$\left. \frac{\partial f(\theta,\tau)}{\partial \tau_i} \right|_{\tau=0} = \tfrac{1}{2} \hbar \varepsilon_{ijk} \theta_j b_k \tag{2.12}$$

where $b_i$ is a real constant dimensionless vector of arbitrary length. Writing $b_i = \lambda a'_i$, where $\lambda = $ const. and $a'_i$ is a unit vector, insertion in (2.7) yields

$$j_i = \frac{\hbar}{2mi}\left(\psi^* \frac{\partial \psi}{\partial q_i} - \frac{\partial \psi^*}{\partial q_i} \psi \right) + \frac{\lambda \hbar}{2m} \varepsilon_{ijk} \frac{\partial \rho}{\partial q_j} a'_k. \tag{2.13}$$

We thus obtain, for all admissible bilinear distributions, just the expression for current given in (1.1), up to the constants $\lambda$ and $a'_i$; varying these parameters generates an infinite set of currents with a 'spin-like' component. As special cases the set includes (1.1) ($\lambda=1, a'_i = a_i$) and (2.8) ($\lambda = 0$).

None of the popular choices for $f$ (as listed in [6,7]) obey (2.12) except for a subset for which $\lambda = 0$. For example, for the Kirkwood distribution ($f = \exp(i\hbar \tau_i \theta_i /2)$) the $f$-term in (2.7) is $(i\hbar/2m)\partial \rho/\partial q_i$ and hence the current is neither real nor conserved (the formalism nevertheless implies the conservation of $\rho$ since the imaginary term cancels with another term in the dynamical equation). An example of a suitable kernel that satisfies (2.2) and implies (2.12) is the function

$$f(\theta,\tau) = \exp\left(\tfrac{1}{2}\hbar\varepsilon_{ijk}\tau_i\theta_j b_k\right). \tag{2.14}$$

However, this function satisfies the conditions assumed in deriving (2.7) through partial integration only for a subset of wavefunctions. To circumvent this restriction we may introduce a Gaussian factor:

$$f(\theta,\tau) = \exp\left(-\tfrac{1}{2}\hbar^2\tau_i\tau_i\theta_j\theta_j\right)\exp\left(\tfrac{1}{2}\hbar\varepsilon_{ijk}\tau_i\theta_j b_k\right). \tag{2.15}$$

A more general option, prompted by an example given by Cohen [5], employs an arbitrary normalized function $\phi$:

$$f(\theta,\tau) = \int \phi\left(q_i - \tfrac{1}{2}ik\hbar\varepsilon_{ikj}\tau_k\theta_j\right)\phi^*\left(q_i + \tfrac{1}{2}ik\hbar\varepsilon_{ikj}\tau_k\theta_j\right)dq \tag{2.16}$$

where $k$ is a real constant with the dimension of length. In this case (2.2) is again obeyed and

$$\left.\frac{\partial f(\theta,\tau)}{\partial \tau_i}\right|_{\tau=0} = -ik\hbar\varepsilon_{ijk}\theta_j \int \phi^*(q)\frac{\partial\phi(q)}{\partial q_k}dq. \tag{2.17}$$

The integral factor in (2.17) is pure imaginary and we choose the parameter dependence of $\phi$ so that this factor equals $ib_k/2k$. Then we obtain (2.13). As an example, we may choose $\phi$ to be a Gaussian,

$$\phi(q) = \left(2\pi\sigma_0^2\right)^{-3/4}\exp\left(-q^2/4\sigma_0^2\right)\exp\left(ib_i q_i/2k\right), \tag{2.18}$$

for which (2.16) reduces to (2.14).

The examples given so far satisfy (2.2b). An example of a bilinear distribution that gives (2.12) but violates this condition is given by

$$f(\theta,\tau) = \exp\left(-\tfrac{1}{2}c\hbar^2\tau_i\tau_i\right)\exp\left(\tfrac{1}{2}\hbar\varepsilon_{ijk}\tau_i\theta_j b_k\right), \; c = \text{const.} \tag{2.19}$$

### 3. Non-bilinear distributions

When $f$ depends on $\psi$, generally no constraint more specific than (2.9) may be stated. We shall show that there are nevertheless kernels within this set that obey (2.12), with (2.2b) valid or not, and hence the spin-like term follows in these cases as well.

As an example of a non-bilinear distribution that obeys (2.2b), we may consider a free system and the kernel (2.16) where $\phi = \psi$, a time-dependent Gaussian.

An example of a useful non-bilinear distribution for which (2.2b) is violated (and which moreover is non-negative) is the following:

$$F(q,p) = \rho(q)\delta\left(p_i - \partial S/\partial q_i - \tfrac{1}{2}\hbar\varepsilon_{ijk}\left(\partial\log\rho/\partial q_j\right)b_k\right) \tag{3.1}$$

where $S$ is the phase of $\psi$. The kernel that generates this distribution is given by

$$f(\theta,\tau) = \frac{\int \rho(q)\exp\left\{i\tau_i\left[\partial S/\partial q_i + \tfrac{1}{2}\hbar\varepsilon_{ijk}\left(\partial\log\rho/\partial q_j\right)b_k\right] + i\theta_i q_i\right\}dq}{\int \exp(i\theta_i u_i)\psi^*\left(u - \tfrac{1}{2}\hbar\tau\right)\psi\left(u + \tfrac{1}{2}\hbar\tau\right)du}. \tag{3.2}$$

The function (3.1) characterizes a generalization of the de Broglie-Bohm phase space (for which $\lambda = 0$ [11]).

## 4. Conclusion

We have examined the role of the phase space approach to non-relativistic quantum mechanics in determining the conserved current associated with the Schrödinger equation. Given the formula (2.6), and for bilinear distribution functions, the dependence of the current on the wavefunction is fixed uniquely, the only freedom being the constant vector $b_i$. An unexpected result is that this spin 0 current simulates the appearance of spin since, apart from a possible deviation of $b_i$ from the spin vector $(\hbar/2)a_i$, it coincides with the current determined uniquely from relativistic considerations. A similar result is obtained for some non-bilinear distributions. It is not clear from the present analysis whether the identity of currents is a coincidence or has a deeper significance.

One of the striking features of (1.1) is that it exhibits 'kinematic interdependence', that is, when $\psi$ factorizes in orthogonal directions the current components are generally coupled [2] (the spin term defines a two-dimensional phase space flow where the Hamiltonian is proportional to $\log\rho$). In contrast to the Wigner function (2.3), for example, which factorizes in this case, the generalized distributions for which $\lambda \neq 0$ represent the interdependence in that they generally do not factorize. This suggests that choosing a distribution function in order to achieve a particular marginal current is an important consideration when formulating a phase space theory. This choice will influence the correspondence rule and local quantities other than current. For example, the expressions for local kinetic energy obtained in our case differ from those studied in [9]. In the case of the kernel (2.14) we obtain

$$\begin{aligned}K(q) &= (1/2m)\int p^2 F(q,p)\,dp \\ &= -\left(\hbar^2/8m\right)\left(\psi^*\nabla^2\psi + \psi\nabla^2\psi^* - 2\nabla\psi^*.\nabla\psi\right) \\ &\quad + \left(i\hbar^2/2m\right)(\nabla\psi^* \times \nabla\psi).b + \left(\hbar^2/8m\right)(b\times\nabla)^2\rho.\end{aligned} \tag{4.1}$$

Evidently, this choice obeys the required condition $\int K\,dq = \langle \hat{p}^2/2m \rangle$ (the proof requires (2.2a) but not (2.2b)). Following the approach of this paper, a further point to consider in assessing admissible expressions for the local kinetic energy is the latter's potential contribution to an energy conservation equation.